\documentclass{appolbnh}
\usepackage{graphicx}
\usepackage{color}
\usepackage{lineno}
\begin{document}

\title{The Roper Resonance $N^*(1440)$ in Nucleon-Nucleon Collisions and the Issue of Dibaryons %
}

\author{Heinz Clement
\thanks{heinz.clement@uni-tuebingen.de}
\address{Physikalisches Institut and Kepler Center for Astro and Particle
  Physics, \\Eberhard-Karls-University of T\"ubingen, \\Auf der Morgenstelle 14,
  D-72076 T\"ubingen, Germany}
}

\maketitle

\begin{abstract}
   In many reactions leading to excitations of the nucleon the Roper resonance
   $N^*(1440)$ can be sensed only very indirectly by complex partial-wave
   analyses. In nucleon-nucleon collisions the  isoscalar single-pion
   production as well as specific two-pion production channels present the
   Roper excitation free of competing resonance processes at a mass of 1370
   MeV and a width of 150 MeV. A detailed analysis points to the formation of
   $N^*(1440)N$ dibaryonic systems during the nucleon-nucleon collision
   process similar to what is known from the $\Delta(1232)N$ threshold. 

\end{abstract}


\section*{Dedication}
This paper is dedicated to the 90th birthday of L. D. Roper, who discovered 
the famous Roper resonance in the nucleon.

\section*{Introduction}

The $N^*(1440)$ resonance has been a puzzle, ever since its discovery in
$\pi N$ phase shifts by L. D. Roper in 1964 \cite{Roper}. In most  respective
investigations no apparent resonance signatures show up directly in the
observables, but have to be revealed by sophisticated partial-wave
analyses. Its resonance parameters show still quite some scatter in its values
\cite{PDG}. Also its nature has been a matter of permanent debate. The finding
that it is in principle of a two-pole structure \cite{Arndt1985} increases its
complexity discussed in many subsequent studies \cite{Cutkowsky,
  Arndt2006,Doring,Suzuki}. Since the quantum numbers of $N^*(1440)$ are
identical to those of the nucleon it also has been associated with the
breathing mode of the nucleon.

Whereas for most resonances the values for the pole position obtained in
partial-wave analyses on the one hand and for the resonance mass and width
obtained by Breit-Wigner fits on the other hand are very close, the situation
is totally different for the Roper resonance. Recent phase-shift analyses of
$\pi N$ and $\gamma N$ data show the pole of the Roper resonance to be about 70 MeV below its canonical value of 1440 MeV. In PDG its pole position is presently quoted to be in the range 1360 - 1380 ($\approx$ 1370) MeV (real part) and 180 - 205 ($\approx$ 190) MeV (2x imaginary part), whereas its Breit-Wigner mass and width are estimated to be in the range m = 1410 - 1470 ($\approx$ 1440) MeV and 250 - 450 ($\approx$ 350) MeV \cite{PDG}.

In nucleon-nucleon and nucleon-nucleus collisions the $N^*(1440)$ excitation
usually sits on top of substantial background, which cannot be removed easily
nor reliably calculated. In the $\alpha p$ experiment at Saclay
\cite{Morsch1992} a bump representing the Roper excitation is observed at m =
1390(20) MeV with $\Gamma = 190(30)$ MeV in the missing mass spectrum,
however, still sitting on large background. Hence the detailed resonance
parameters depend substantially on the treatment of the background as
demonstrated, $\it{e.g.}$, in Ref. \cite{Oset1996}. Analyses of high-energy $pp$
scattering give similar values for the Roper excitation \cite{Morsch2005}.

\section*{Isoscalar Single-Pion Production in $NN$ Collisions}

The beauty of isoscalar single-pion production is that the usually
overwhelming $\Delta$ excitation is eliminated by isospin selection in this
process. Hence the excitation of the next higher-lying resonance, the Roper
resonance, can be observed free of any resonance background. The total cross
section of isoscalar single-pion production can be obtained from that of the
purely isovector reaction $pp \to pp\pi^0$ and that of the isospin-mixed
reaction $pn \to pp\pi^-$ by the isospin relation:

\begin{eqnarray}
  \sigma_{pn \to NN\pi}(I=0) = 3~(\sigma_{pn \to pp\pi^-} - 1/2~\sigma_{pp \to pp\pi^0})
\end{eqnarray}

Both reactions have been measured exclusively and kinematically complete (with
overconstraints) by WASA-at-COSY in the energy range $T_p =$ 1.0 - 1.35 GeV
($\sqrt s =$ 2.3 - 2.45 GeV) \cite{IsoscalarNNpi,PRC106(2022)065204}. Fig. 1
shows the energy dependence of the total cross section of the $pn \to pp\pi^-$
reaction from threshold up to $\sqrt s =$ 2.5 GeV. Plotted are all available
data from previous measurements together with the WASA results, which are
given by the full (red) dots. 

\begin{figure}[htb!]
\centering
{
    \includegraphics[width=0.7\textwidth,keepaspectratio]{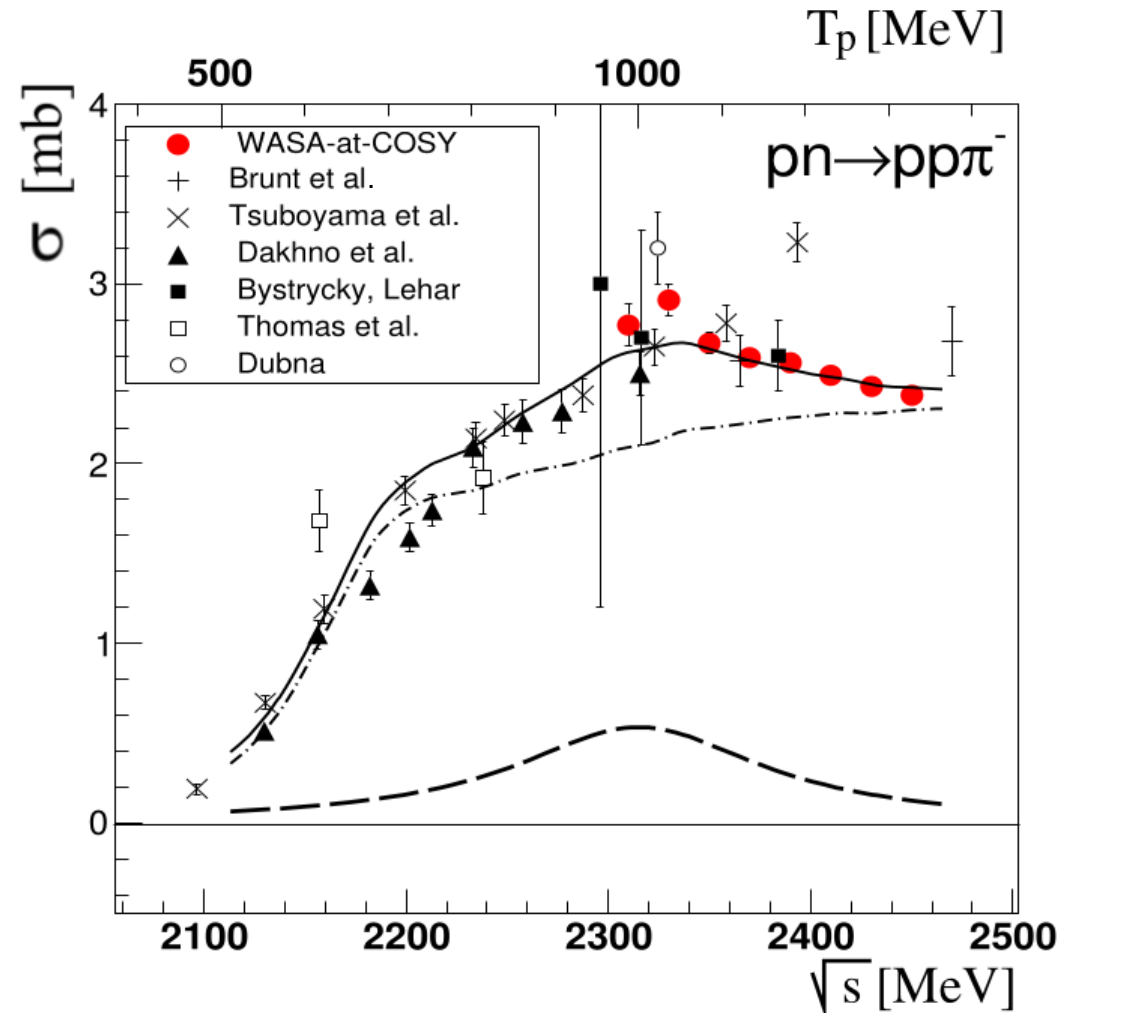} 
}

\centerline{\parbox{1\textwidth}{
\caption[] {\protect\small
Energy dependence of the total cross section  for the $pn \to pp\pi^-$
reaction. Full (red) dots represent results from WASA-at-COSY
\cite{IsoscalarNNpi}, other symbols denote results from earlier work
\cite{Bystricki,Tsuboyama,Dakhno,Brunt,Thomas,Abdivaliev}. The dash-dotted
line shows the isovector contribution. The dashed curve at the bottom gives a
Breit-Wigner fit to the difference between dash-dotted curve and the data  for
the $pn \to pp\pi^-$ reaction. The solid line gives the sum of dashed and
dash-dotted curves. From Ref. \cite{PRC106(2022)065204}.
} 
\label{fig:Fig1} } }
\end{figure}

Starting from threshold we observe a strongly rising cross section, which may
be associated with $t$- and $s$-channel $\Delta$ excitation
\cite{PRC106(2022)065204}. Around $\sqrt s =$ 2.2. GeV the cross section
levels off somewhat before starting rising again towards higher energies. Such
a rise is expected from the excitation of the Roper resonance. However, what
is unexpected, is that the cross section starts falling again beyond $\sqrt s
=$ 2.3 GeV leading thus to a pronounced bump structure in the total cross
section. 

The dash-dotted curve shown in Fig. 1 represents a fit to
corresponding data for the $pp \to pp\pi^0$ reaction (see Fig. 3 in
Ref. \cite{PRC106(2022)065204}) and gives its total cross section divided by
two, $\it{i. e.}$, it represents just the isovector contribution to the $pn
\to pp\pi^-$ channel.
The difference of the dash-dotted 
curve to the total cross section of the $pn \to pp\pi^-$ reaction can be well
represented by a Breit-Wigner curve with m = 2310 MeV and $\Gamma =$ 150 MeV
plotted by the dashed curve at the bottom of Fig. 1. According to eq. (1) this
dashed curve represents now the isoscalar cross section of single-pion
production divided by three. The full curve in Fig. 1 is just the sum of dashed
and dash-dotted curves reproducing the data of the $pn \to pp\pi^-$ reaction
reasonably well. 

Fig. 2 displays the deduced isoscalar single-pion production cross section
together with all available previous data
\cite{Tsuboyama,Dakhno,Rappenecker,Sarantsev2004}, in particular also with the
results of the partial-wave analyses of Ref. \cite{Sarantsev}, which are
plotted in Fig. 2 by open crosses surrounded by a hatched band indicating the
uncertainties. As in Fig. 1 a Breit-Wigner fit is displayed with m = 
2310 MeV and $\Gamma =$ 150 MeV.  For comparison we also show the expected
energy dependence of a conventional $t$-channel excitation of the Roper
resonance with subsequent $p$-wave pion emission \cite{IsoscalarNNpi}
arbitrarily normalized to the data point at $\sqrt s =$ 2260 MeV. Due to the
strong energy dependence of the $p$-wave emission we would have expected a
steeply increasing cross section -- similar to what is observed for the
$\Delta$ excitation. 

\begin{figure}[htb!]
\centering
{
    \includegraphics[width=0.7\textwidth,keepaspectratio]{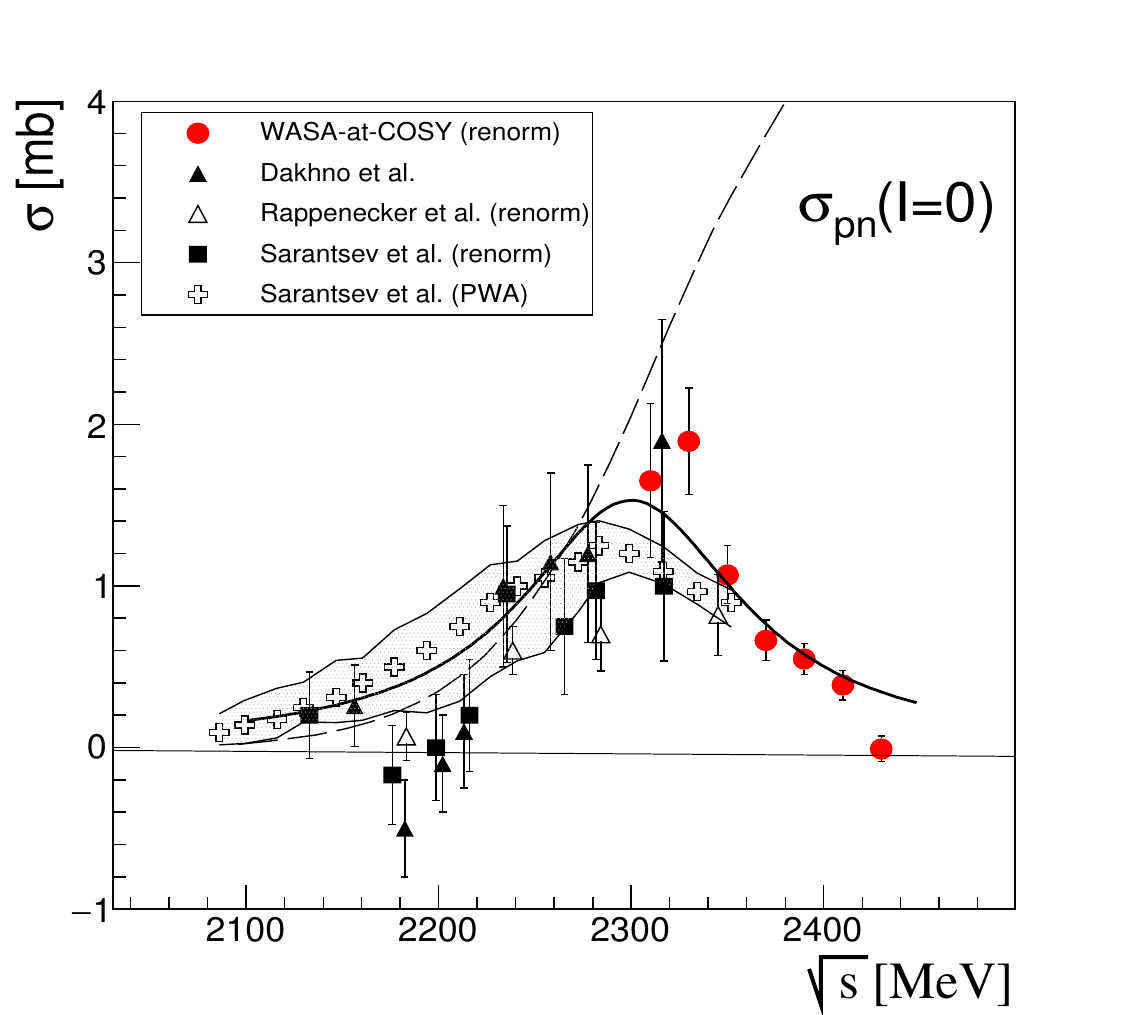} 
}

\centerline{\parbox{1\textwidth}{
\caption[] {\protect\small
The $pn$-induced isoscalar single-pion production total cross section in
dependence of the total c.m. energy $\sqrt s$. Shown are the results from
WASA-at-COSY \cite{IsoscalarNNpi,PLB806} and
Refs. \cite{Tsuboyama,Dakhno,Rappenecker,Sarantsev2004} as well as the results
of the partial-wave analyses of Ref. \cite{Sarantsev} (open crosses surrounded
by a hatched band, which indicates the uncertainties). The solid line
represents a Breit-Wigner with m = 2310 MeV and $\Gamma =$ 150 MeV. The dashed
line gives the expected energy dependence of a $t$-channel Roper excitation
\cite{IsoscalarNNpi} adjusted arbitrarily in height to the data point at $\sqrt
s =$ 2260 MeV. From Ref. \cite{PRC106(2022)065204}.
} 
\label{fig:Fig2} } }
\end{figure}

The resonance-like structure in the region of the Roper excitation points to
the formation of a $N^*(1440)N$ dibaryonic system. To investigate, whether
this is, indeed, connected with the excitation of the Roper resonance, we plot
in Fig. 3 the isoscalar $N\pi$ invariant mass distribution $M_{N\pi}(I=0)$
obtained from the WASA measurement. As we can see, there is a pronounced bump
above practically no background in the region of the Roper resonance, which
again can be well represented by a Breit-Wigner with m = 1370 MeV and $\Gamma
=$ 150 MeV.

\begin{figure}[htb!]
\centering
{
    \includegraphics[width=0.7\textwidth,keepaspectratio]{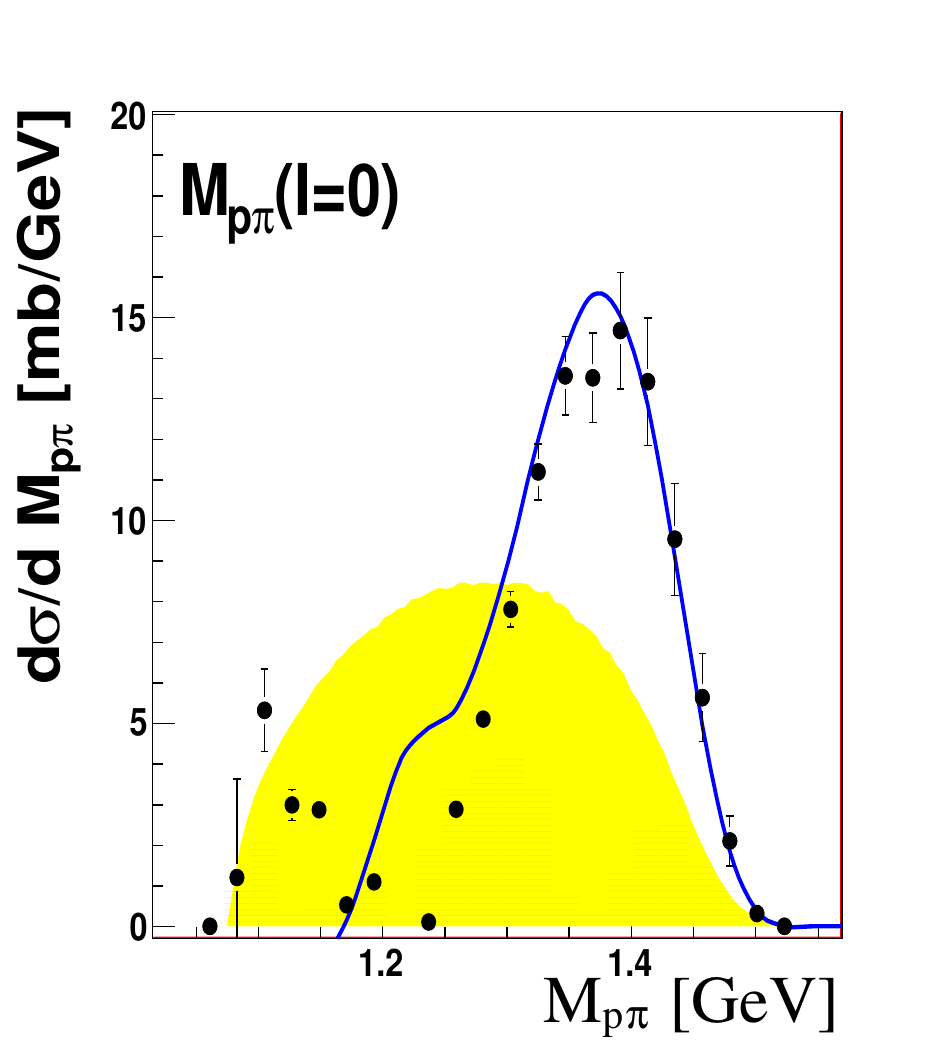} 
}

\centerline{\parbox{1\textwidth}{
\caption[] {\protect\small
The isoscalar $N\pi$ invariant mass distribution $M_{N\pi}(I=0)$ as obtained
from the WASA measurements of the $pp \to pp\pi^0$ and $pn\to pp\pi^-$
reactions. The yellow area represents a pure phase space distribution, the
solid line a $t$-channel calculation for the Roper excitation with m = 1370
MeV and $\Gamma =$ 150 MeV. From Ref. \cite{IsoscalarNNpi}.
} 
\label{fig:Fig3} } }
\end{figure}

Adding the mass of a nucleon to the Roper mass extracted from Fig. 3, then we
end up with 2310 MeV, which is just the mass of the bump structure seen in the
isoscalar total cross section. If we relate the effective Roper mass seen here
to its pole position, then we observe the $N^*(1440)N$ system just at
threshold. If we relate the effective Roper mass seen here with its nominal
Breit-Wigner mass of 1440 MeV, then the Roper appears to be bound by about 70
MeV in the $N^*(1440)N$ system and its observed reduced width can be well
understood by the momentum-dependence of its $p$-wave decay. 

The width of the Roper resonance observed in the isoscalar $N\pi$
invariant-mass spectrum equals that obtained for the bump structure in the
isoscalar total cross section. This is actually not very surprising. As we
know, e.g., from tetra- and pentaquark studies \cite{PDG}, due to the tiny
available phase space near threshold the decay width of near-threshold states
is tiny, if the decay products are hadronically stable. In  our case the decay
products of the $N^*(1440)N$ dibaryonic system are $N$ and $N^*(1440)$ and the
latter is not hadronically stable and has a large width. Hence we see just the
width of the Roper resonance in the dibayonic system. 

Next we have to consider spin and parity of the $N^*(1440)N$ system. From the
partial-wave analyses of Sarantsev \textit{et al.} \cite{Sarantsev} we know
that there 
are two dominating isoscalar $NN$ partial waves in the region of interest: the
$^3S_1-^3D_1$ $pn$ partial wave, where $N$ and $N^*(1440)$ are in relative $S$ wave
leading to $I(J^P) = 0(1^+)$ and the $^1P_1$ $np$ partial wave, where $N$ and
$N^*(1440)$ are in relative $P$ wave yielding $I(J^P) = 0(1^-)$. Note that
$1^+$ and $1^-$ are the only possible $J^P$ combinations for isoscalar $S$ and
$P$ waves in the $NN$ system. 

At a first glance it might not appear very convincing that just two resonances
sit practically on top of each other producing that way a single
resonance-like structure in the total cross section. However, exactly such a
scenario is observed also near the $\Delta N$ threshold, where the isovector
$0^-, 2^+, 2^-$ and $3^-$ dibaryonic states happen to have similar masses and
widths with differences small compared to their width \cite{Igor,Oh,ANKE}. For
recent reviews about this issue see, \textit{e.g.}, Refs. \cite{PPNP,CPC}. And 
since the width of the $N^*(1440)N$ states is still substantially larger than
that of the $\Delta N$ states, small differences in mass and width are washed
out in the summed up shape. 

\section*{Isoscalar Two-Pion Production in $NN$ Collisions}

In addition to its single-pion decay the Roper resonance decays also by
two-pion emission, though it is not the dominating decay process. Hence the
isoscalar $N^*(1440)N$ system should show up also in isoscalar two-pion
production -- if the background situation is favorable. Indeed, there is such a
situation in the $pn \to d\pi^0\pi^0$ reaction. The energy dependence of its
total cross section measured by WASA \cite{PRL106,PLB721} is shown in
Fig. 4.

\begin{figure}[htb!]
\centering
{
    \includegraphics[width=0.7\textwidth,keepaspectratio]{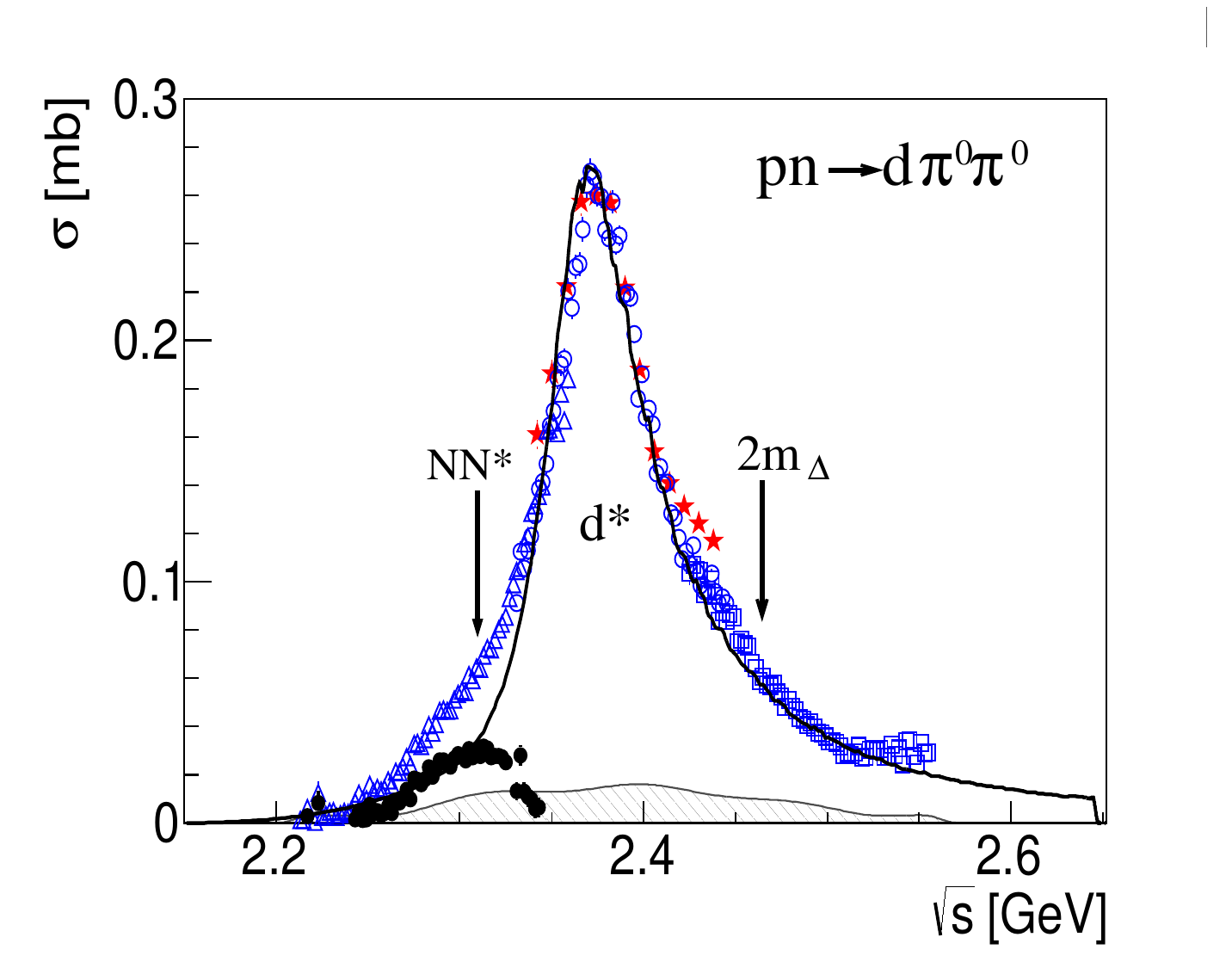} 
}

\centerline{\parbox{1\textwidth}{
\caption[] {\protect\small
The energy dependence of the total cross section of the $pn \to d\pi^0\pi^0$
reaction measured by WASA-at-COSY \cite{PRL106,PLB721} (blue and red
symbols). The hatched area represents an estimate of systematic
uncertainties. The solid curve gives a calculation of the $d^*(2380)$ dibaryon
resonance with momentum-dependent widths \cite{NPA2017} including both Roper
and $\Delta\Delta$ $t$-channel excitations as background reactions. The black
solid dots represent the difference between this calculation and the data at
low energies.
  From Ref. \cite{IsoscalarNNpi}.
} 
\label{fig:Fig4} } }
\end{figure}

The most dominating 
feature there is, of course, the excitation of the $d^*(2380)$ dibaryon
resonance. The solid curve gives a description to this resonance with a momentum
dependent width \cite{NPA2017}, which describes the data very well except in
its low-energy tail. There the resonance curve underpredicts the data
substantially. Actually, this is just the location, where we expect strength
due to the two-pion decay of the $N^*(1440)N$ system. If we substract the 
curve from the data in this region, then we obtain some bell-shaped distribution
around $\sqrt s \approx$ 2.3 GeV (black dots), the high-energy side of which
is, of course, highly dependent on the $d^*(2380)$ description. It is
intriguing to associate this bump with the two-pion decay of the $N^*(1440)N$
systems. Having here a peak cross section of about 30 $\mu b$ we may
estimate the contributions from other two-pion production channels with
isoscalar contributions by isospin rules and end up finally with a total of
roughly 150 $\mu b$ \cite{PRC106(2022)065204}.

We note in passing that the two-pion production via the $N^*(1440)N$ systems
can be described also successfully as sequential single-pion processes by
accounting for the two-pion decay of the Roper resonance. Using the formalism
of Oset \textit{et al.} \cite{Osetseq} for sequential single-pion production
we arrive at the proper value for the observed cross section \cite{seq}. 

\section*{Branching Ratios of the $N^*(1440)N$ Dibaryonic Systems}

Having identified all inelastic decay channels we may extract the branching
ratios for the transitions $N^*(1440)N \to NN, NN\pi,$ and $NN\pi\pi$ in
ana-logy to what was done for $d^*(2380)$ \cite{EPJA51(2015)87}. From the
partial-wave analyses of Ref. \cite{Sarantsev} we infer that about 25$\%$ of
the peak cross section in isoscalar single-pion production is due to the
$I(J^P) = 0(1^+)$ state and 75$\%$ due to the $I(J^P) = 0(1^-)$ state. By
using unitarity we arrive at elastic branchings of 0.04 and 0.15,
respectively. Similarly the branching into the $NN\pi$ and $NN\pi\pi$ channels
are roughly 0.8 and 0.2, respectively  \cite{PRC106(2022)065204}. This means,
these resonances reside predominantly in the inelastic channels, in particular
in the $NN\pi$ channel. Its tiny elasticities make it very hard to sense them
in elastic scattering. 

\section*{Isovector Two-Pion Production in $NN$ Collisions}

Since the $\Delta$ resonance decays only by single-pion emission, even the
isovector two-pion production is free of single-$\Delta$ contributions. Only
above $\sqrt s =$ 2.3 GeV $\Delta$ degrees of freedom enter by the
$\Delta\Delta$ 
excitation process. Hence at lower energies the Roper excitation constitutes
the only resonance reaction. The situation is particularly attractive in the
$pp \to pp\pi^0\pi^0$ channel due to its reduced isospin combinations in its
subsystems \cite{EPJA35(2008)317}. Exclusive and kinematically complete
measurements have been carried out by PROMICE/WASA and CELSIUS/WASA in this
particular region. From the different invariant-mass and opening angle spectra
the decay routes $N^*(1440) \to \Delta\pi \to N\pi\pi$ and $N^*(1440) \to
N\sigma$ could be well separated and their relative branching determined. In
contrast to previous listings \cite{PDG2006} we find a ratio of 1.0 (1) for
these two branchings in good agreement with more recent evaluations
\cite{PDG}. 

In Fig. 5 we show the measured energy dependence of the total cross
section. After a steep rise at threshold the cross section levels off near
$\sqrt s =$ 2.3 GeV before it starts rising again beyond $\sqrt s =$ 2.4
GeV. Whereas the first rise is due to Roper excitation, the second rise is
associated with $\Delta\Delta$ excitation. If we make a isospin decomposition
of all two-pion production channels with regard to these excitation processes,
then we get an energy dependence for $N^*$ excitations, which is given by
asterisk symbols in Fig. 5 \cite{PLB679(2009)30}. This distribution shows
again a bump-like structure peaking around 2.3 GeV and a width of 140
MeV. This suggests that also here we deal possibly with a $N^*(1440)N$ system,
but now with $I(J^P) = 1(0^+)$ connected with the $^1S_0$ $NN$ partial wave in
the initial $pp$ system.

\begin{figure}[htb!]
\centering
{
    \includegraphics[width=0.7\textwidth,keepaspectratio]{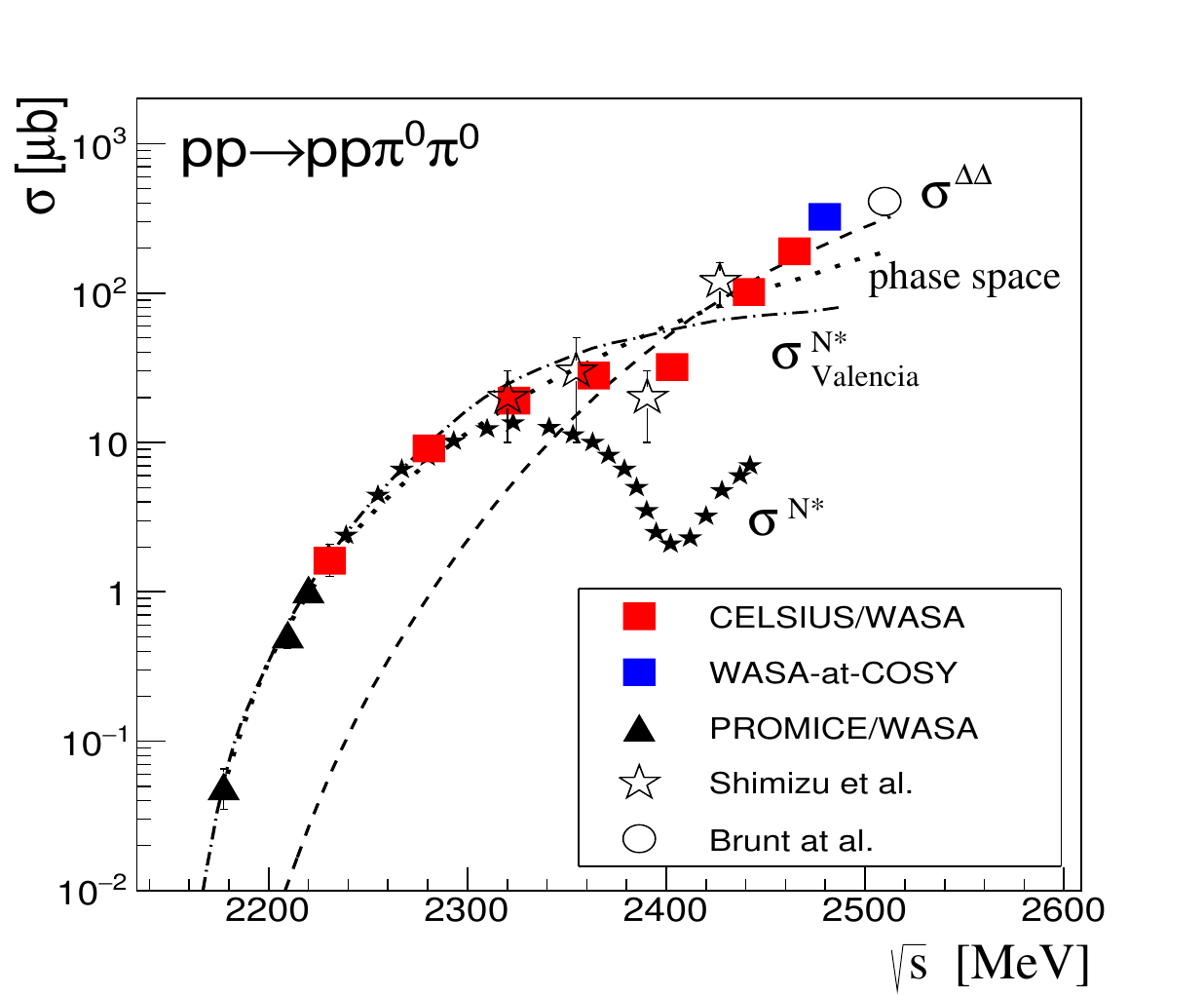} 
}

\centerline{\parbox{1\textwidth}{
\caption[] {\protect\small
The energy dependence of the total cross section of the $pp \to pp\pi^0\pi^0$
reaction measured by PROMICE/WASA \cite{PROMICE}, CELSIUS/WASA
\cite{PLB679(2009)30}, 
WASA-at-COSY \cite{PLB706} and earlier measurements \cite{Brunt,Shimizu}.The
dotted, dash-dotted and dashed lines give the expected energy dependence for
pure phase space as well as  $t$-channel Roper and $\Delta\Delta$ excitations,
respectively \cite{Luis}. The filled stars display the result of an isospin
analysis of all two-pion production channels regarding $N^*$ excitations
\cite{PLB679(2009)30}.
  From Ref. \cite{Kukulin2020}.
} 
\label{fig:Fig5} } }
\end{figure}

\section*{Influence of the $N^*(1440)N$ Dibaryonic Systems on the $NN$ Interaction}

Dibaryons constitute hexaquark configurations and hence are a link between
meson-nucleon and quark degrees of freedom in the interaction of two
nucleons. These considerations have been taken into account in the dibaryon $NN$
interaction model of Kukulin, Platonova $\it{et~al.}$
\cite{Kukulin2022,Kukulin2020} based 
on ideas developed in Refs. \cite{Kukulin2001,Kukulin2002}. In this model the
intermediate and short range part of the $NN$ interaction, which covers the
region, where the nucleons overlap, is described by $s$-channel exchange of
intermediate dibaryons representing six-quark configurations. That way it
replaces the $t$-channel exchange of a multitude of heavy mesons in
conventional models. The only meson exchange kept in this model is the
one-pion exchange for the long range part of the interaction, where the two
nucleons practically no longer overlap. The beauty of this model is that it
works even 
above the pion emission threshold, since the intermediate dibaryons get then
on mass shell and decay by meson emission. That way this model automatically
includes inelastic scattering and hence works up to the GeV range. In addition
and most important, the intermediate dibaryons needed in this model for the
description of the low angular momentum partial waves can be cross checked
experimentally by searching for them in meson-production measurements. 

In Ref. \cite{Kukulin2022} it is demonstrated that this model can describe
very successfully the experimental phase shifts \cite{SAID2016,SAID2020} both in
the real and in the imaginary parts up to the GeV range for $S,P,D$ and $F$
waves. The dibaryons extracted in their model turn out to be in very good
agreement with the
experimental findings. In Ref. \cite{Kukulin2020} it is shown that the looping
of the $N^*(1440)N$ configurations with $I(J^P) = 1(0^+)$ and $0(1^+)$ in the
Argand diagrams of $^1S_0$ and $^3S_1$ partial wave is very small due to their
tiny elasticities and hence it is hard to identify them uniquely in
partial-wave analyses. Nevertheless, as demonstrated in
Ref. \cite{Kukulin2020} the influence of $N^*(1440)N$ resonances on the phase
shifts turns out to be crucial over the full energy range -- in particular for
the $S$ waves (Fig. 6), where the overlap of the two nucleons is at maximum.

\begin{figure}[htb!]
\centering
{
    \includegraphics[width=1.0\textwidth,keepaspectratio]{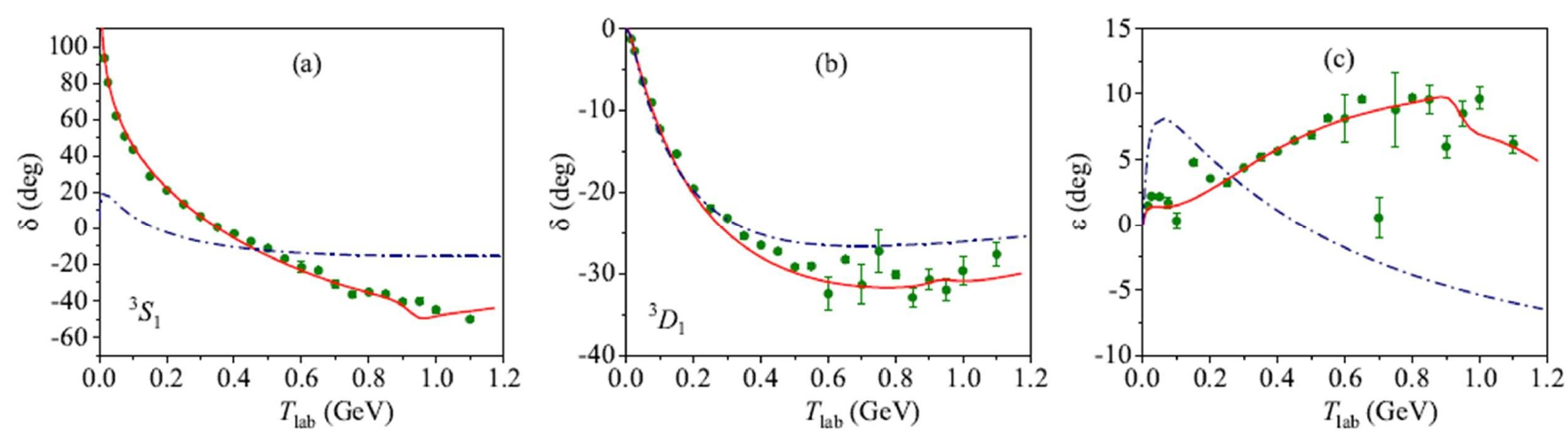} 
}

\centerline{\parbox{1\textwidth}{
\caption[] {\protect\small
Phase shifts $\delta$ for the coupled $NN$ partial waves $^3S_1$ (a) and
$^3D_1$ (b) as well as the mixing angle $\epsilon$. The solid dots display the
single-energy solutions of the SAID partial-wave analyses \cite{SAID2016}, the
solid curves show the results of the dibaryon $NN$ interaction and the
dash-dotted lines the results of the pure one-pion exchange.
  From Ref. \cite{Kukulin2020}.
} 
\label{fig:Fig6} } }
\end{figure}

\section*{Conclusions}

Even 60 years after its discovery by L.D. Roper the $N^*(1440)$ resonance is
still a matter of utmost attraction both theoretically and experimentally. The
fact that its pole values differ strongly from its Breit-Wigner mass and width
increases its complexity and has caused in the past quite some confusion. In
$NN$ collisions the Roper excitation appears usually as a bump above
substantial background and seems to have lower values for its mass and width
as compared to what is observed in $N\pi$ and $N\gamma$ studies. 

In exclusive and kinematically complete (with overconstraints) measurements at
CELSIUS/WASA and WASA-at-COSY the Roper resonance has been observed free of
background in single- and double-pion production, where it exhibits a
resonance-like energy dependence in total cross sections. This is shown as
being due to the formation of $N^*(1440)N$ dibaryonic systems, where the Roper
resonance is bound by roughly 70 MeV. This binding explains also its reduced
width of 150 MeV observed in the invariant mass spectrum $M_{N\pi}(I=0)$.

That way the puzzling discrepancy between $N\pi$ and $N\gamma$ results for the
Roper resonance and those obtained by $NN$ collisions is resolved by the
finding that in the latter case the Roper resonance merges into a $N^*(1440)N$
configuration. 

The observation of $N^*(1440)N$ dibaryons is very important for the
understanding of the $NN$ interaction, where they turn out to be crucial for
the understanding of the $NN$ partial waves, in particular the $S$ and $P$
waves, where the colliding nucleons strongly overlap and hence reveal their
quark degrees of freedom.

\section*{Acknowledgments}

This work was supported by BMBF and DFG. I acknowledge valuable discussions
with L. Alvarez-Ruso, M. Bashkanov, Y. Dong, E. Doroshkevich, A. Gal,
Ch. Hanhart, V. Kukulin (deceased), E. Oset, M. Platonova, M. Schepkin
(deceased), 
T. Skorodko, I. I. Strakovsky, C. Wilkin and Z. Zhang.


\end{document}